\documentclass{pasj01}
\Received{$\langle$reception date$\rangle$}
\Accepted{$\langle$acception date$\rangle$}
\Published{$\langle$publication date$\rangle$}

\begin{document}

\title{Spatially resolved spectroscopy of nonthermal X-rays in RX~J1713.7$-$3946 with Chandra}

\author{Tomoyuki~\textsc{Okuno},\altaffilmark{1,}$^{*}$
 Takaaki~\textsc{Tanaka}\altaffilmark{1},
 Hiroyuki~\textsc{Uchida}\altaffilmark{1},
 Hideaki~\textsc{Matsumura}\altaffilmark{1,2},
 and Takeshi~Go \textsc{Tsuru}\altaffilmark{1}}
\altaffiltext{1}{Department of Physics, Kyoto University, Kitashirakawa Oiwake-cho, Sakyo, Kyoto 606-8502, Japan}
\altaffiltext{2}{Kavli Institute for the Physics and Mathematics of the Universe (WPI), The University of Tokyo Institutes for Advanced Study, The University of Tokyo, 5-1-5 Kashiwanoha, Kashiwa, Chiba 277-8583, Japan }

\email{okuno.tomoyuki.32r@kyoto-u.jp}


\maketitle

\begin{abstract}
The young shell-type supernova remnant (SNR) RX~J1713.7$-$3946 has been studied as a suitable target to test the SNR paradigm for the origin of Galactic cosmic rays. 
We present a spatially resolved spectroscopy of the nonthermal X-ray emission in RX~J1713.7$-$3946 with Chandra.
In order to obtain X-ray properties of the filamentary structures and their surrounding regions,
we divide the southeastern (SE), southwestern (SW), and northwestern (NW) parts of the SNR into subregions on the typical order of several $10\arcsec$ and extract spectra from each subregion.
Their photon indices are significantly different among the subregions with a range of $1.8 < \Gamma <  3$.
In the SE part, the clear filaments are harder ($\Gamma \sim 2.0$) than the surrounding regions.
This is a common feature often observed in young SNRs and naturally interpreted as a consequence of synchrotron cooling.
On the other hand, the bright filamentary regions do not necessarily coincide with the hardest regions in the SW and NW parts.
We also find the SW filamentary region is rather relatively soft ($\Gamma \sim  2.7$).
In addition, we find that hard regions with photon indices of 2.0--2.2 exist around the bright emission although they lie in the downstream region and does not appear to be the blast wave shock front. 
Both two aforementioned characteristic regions in SW are located close to peaks of the interstellar gas.
We discuss possible origins of the spatial variation of the photon indices, paying particular attention to the shock-cloud interactions. 
\end{abstract}

\section{Introduction}
RX~J1713.7$-$3946 (a.k.a. G347.3$-$0.5) is a shell-type supernova remnant (SNR) discovered on the Galactic plane in the ROSAT All-Sky Survey \citep{Pfeffermann1996}.
The age and the distance of RX~J1713.7$-$3946 are estimated to be $\sim 2000~\mathrm{yr}$ \citep{Wang1997} and  $\sim 1~\mathrm{kpc}$ \citep{Koyama1997,Fukui2003}, respectively.
Because of the detection of synchrotron X-ray (e.g., \cite{Koyama1997}) and $\gamma$-ray emissions (e.g., \cite{Abdo2011,HESS2016}), RX~J1713.7$-$3946 has been paid attention as an accelerator of cosmic ray (CR) particles in the Galaxy.
Both the synchrotron X-rays and $\gamma$-rays have relatively hard spectra compared to other SNRs. Thus, one of the key parameters regarding RX~J1713.7$-$3946 is 
the maximum acceleration energy achieved at its expanding shock (e.g., \cite{Uchiyama2007}), which is closely related to the fundamental question on CRs: whether or not SNRs are acceleration sites of CRs up to the knee in the CR spectrum, or the energy of $\sim 3 \times 10^{15}~{\rm eV}$. 

X-ray observations revealed that the nonthermal emission is dominant almost in the entire part of the SNR (e.g., \cite{Slane1999,Tanaka2008}), although thermal emission, most probably from shocked ejecta, was recently detected in the central region \citep{Katsuda2015}.
The complex filamentary structure was found in the northwestern (NW) part with a superb angular resolution ($\sim 0.5\arcsec$) of Chandra \citep{Uchiyama2003,Lazendic2004}. 
\citet{Uchiyama2007} discovered year-scale time variability of small knots in the filaments. 
The authors argued that a mG-scale magnetic field is necessary to account for the variability, and that their result provided evidence 
of the magnetic field amplification at the shock as predicted by theoretical works (e.g., \cite{Bell2001,Bell2004}).

RX~J1713.7$-$3946 is known to be interacting with atomic and molecular gas (e.g., \authorcite{Fukui2003}~\yearcite{Fukui2003},~\yearcite{Fukui2012}).
\citet{Fukui2012} pointed out that the distribution of the $\gamma$-rays correlates with that of the interstellar medium (ISM).
The spatial correlation between the X-rays and the ISM was also reported by \authorcite{Sano2013}~(\yearcite{Sano2013}; \yearcite{Sano2015}).
In particular, \citet{Sano2015} comprehensively compared to the distributions of X-ray spectral parameters with those of the ISM, 
and discussed possible effects of shock-cloud interactions on particle acceleration through, for example, magnetic field amplification. 
RX~J1713.7$-$3946 is a good target to investigate such effects because of the clear shock-cloud interactions.
However, the comparison was performed with angular scales of 2\arcmin--8\arcmin~mainly due to the moderate angular resolution of Suzaku, 
and thus the authors were not able to probe spectral properties of each of the aforementioned filamentary structures. 
The hardness-ratio map by \citet{Uchiyama2003} in fact hints small changes of spectral parameters in an angular scale smaller than 1\arcmin. 
Also, according to the three-dimensional magnetohydrodynamic simulations, \citet{Inoue2012} predicted that shock-cloud interactions result in 
magnetic field and gas density structures in a physical scale  of $\sim 0.05$~pc, which corresponds to $\sim 10\arcsec$ in the case of RX~J1713.7$-$3946
presumably located at 1~kpc. 

Here we present a spatially resolved spectroscopy of RX~J1713.7$-$3946 with Chandra, 
aiming to study X-ray properties of the filamentary structures and their surrounding regions.
In section \ref{observation}, we describe the selection of the datasets and their data reduction.
Section \ref{analysisresults} presents our analysis process and the spatial distributions of the X-ray emission properties, and then the results are discussed in section \ref{discussion}.
Finally, we summarize our conclusions in section \ref{conclusions}.
In this paper, errors in the text and error bars in the figures indicate 90\% and 1$\sigma$ confidence intervals, respectively.

\begin{figure}
 \begin{center}
  \includegraphics[width=80mm]{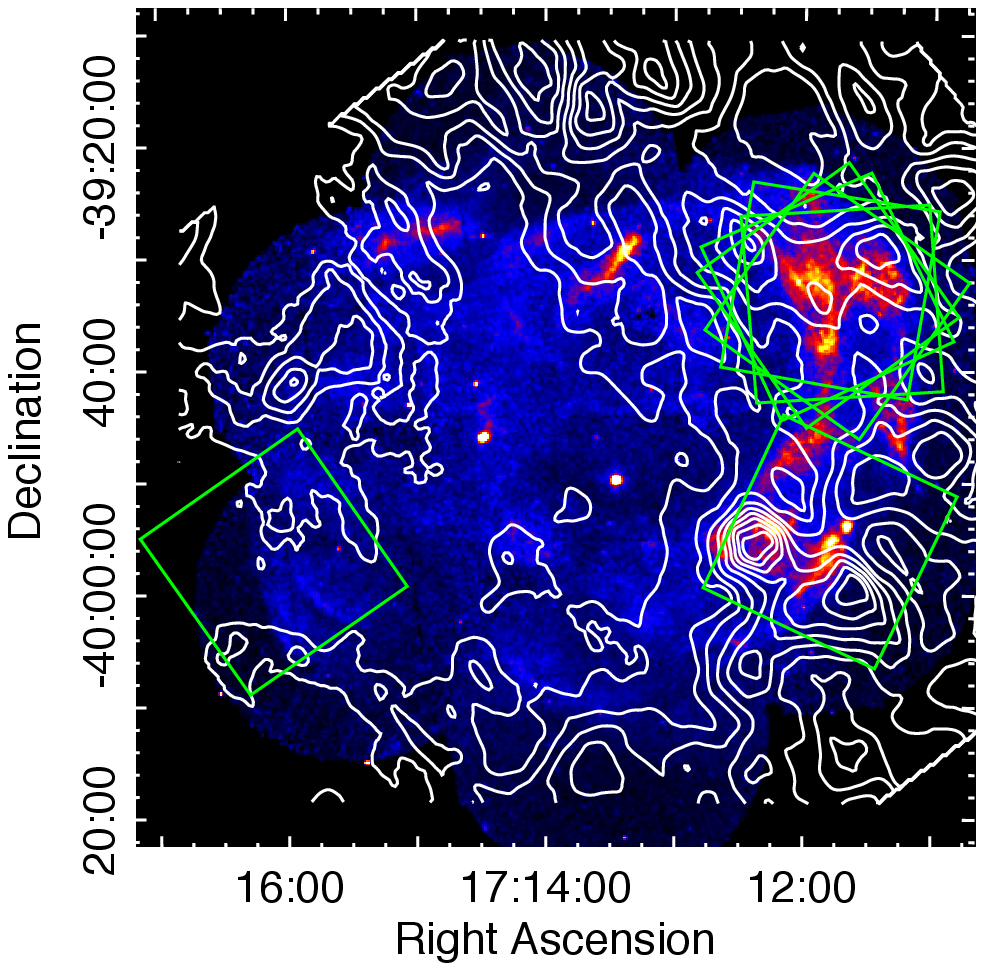}
 \end{center}
 \caption{X-ray (0.5--8~keV) image of RX~J1713.7$-$3946 obtained with XMM-Newton (K.~Mori 2017 private communication). The color scale is in linear. The overlaid white contours represent the proton column density estimated from $^{12}$CO($J = 1\textrm{--}0$) and H\emissiontype{I} data observed with NANTEN, and ATCA \& Parkes telescopes, respectively \citep{Sano2015,Moriguchi2005,McClure2005}. The seven green squares represent the ACIS-I FOV of the observations in table~\ref{observations}.}\label{wholeimage}
\end{figure}

\begin{table}
 \tbl{RX~J1713.7$-$3946 observation log.}{
 \begin{tabular}{rcccc} \hline
   ObsID  & Region  & Start date & Effective exposure\\ && &(ks) \\ \hline 
    6370 & NW & 2006-05-03 & 29.8 \\
    10090  &NW& 2009-01-30 & 28.4\\ 
    10091 & NW & 2009-05-16 & 29.7 \\
    10092 & NW & 2009-09-10 & 29.2 \\
    12671 & NW & 2011-07-01 & 89.9 \\
    5561 & SW & 2005-07-09 & 29.0 \\
    10697 & SE & 2009-05-15 & 57.4 \\ \hline
    \end{tabular}}
    \label{observations}
    \begin{tabnote}
    \end{tabnote}
\end{table}

\begin{figure*}
 \begin{center}
  \includegraphics[width=160mm]{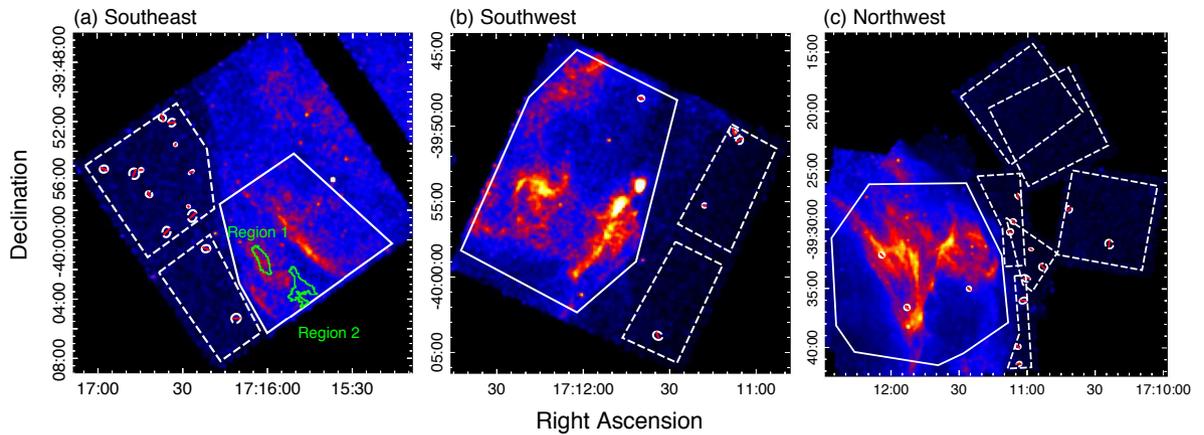}
 \end{center}
 \caption{(a) SE exposure corrected X-ray (0.7--7~keV) image scaled in range of $(0\textrm{--}1) \times 10^{-6} \mathrm{ph~s^{-1}~cm^{-2}}$. (b) SW exposure corrected X-ray (0.7--7~keV) image scaled in range of $(0\textrm{--}2) \times 10^{-6} \mathrm{ph~s^{-1}~cm^{-2}}$. (c) NW exposure corrected X-ray (0.7--7~keV) mosaic image scaled in range of $(0\textrm{--}2) \times 10^{-6} \mathrm{ph~s^{-1}~cm^{-2}}$. The color scale of (a)-(c) is in linear. White solid and dash lines represent source and background regions, respectively. The regions drawn red diagonal lines were excluded from our analysis. Two regions in (a) surrounded by green solid lines (regions 1 and 2) were used for extracting the example spectra shown in figure~\ref{examplespec}.}\label{expcorimg}
\end{figure*}

\begin{figure}
 \begin{center}
  \includegraphics[width=80mm]{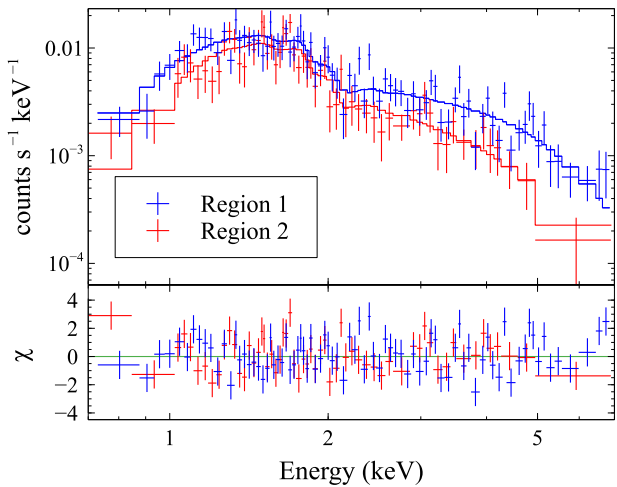}
 \end{center}
 \caption{Hardest spectrum (extracted from region 1) and softest one (extracted from region 2) indicated as blue and red, respectively.
Their best-fit models are indicated as solid lines. 
Residuals are shown in the lower panel.}\label{examplespec}
\end{figure}

\section{Observations and data reduction}\label{observation}
RX~J1713.7$-$3946 was observed several times from 2000 to 2011 with the Advanced CCD Imaging Spectrometer (ACIS)\footnote{$\langle$http://cxc.harvard.edu/cal/Acis$\rangle$.} aboard Chandra. 
All the source spectra we analyzed were extracted from ACIS-I, which is an array of four front-illuminated (FI) chips.
We also used FI chips of ACIS-S for background estimation, and did not use back-illuminated chips because of the different instrumental background levels. 

Some datasets were excluded from our analysis. 
As we detail in section \ref{analysis}, we perform spatially resolved spectroscopy with an angular scale of several $10\arcsec$. 
The shell expansion would smear out position-dependent spectral changes if the X-ray emitting structures move $\sim 10\arcsec$ or more due to the proper motion. 
The proper motions in the NW shell and the southeastern (SE) rim were both reported to be $\sim$ 0.8~arcsec~yr$^{-1}$ \citep{Tsuji2016,Acero2017}.
Therefore, the proper motion between two epochs about 10 years apart may affect our spectral analysis, and thus we did not analyze data from the two observations in 2000 (ObsID 736 and 737). 
In addition to the proper motions, we also considered the fields of view (FOV) when choosing observations.
Since the SNR entirely occupies the FOV of the observation in 2005 (ObsID 5560), we were not able to extract off-source spectra for backgrounds.
Therefore, this observation was also excluded from our analysis.

The observations we used for the analysis described below are summarized in table~\ref{observations}, and their ACIS-I FOVs are indicated in figure~\ref{wholeimage}. 
We reprocessed the data using \texttt{chandra\_repro} task in the analysis software for Chandra, CIAO version 4.9\footnote{$\langle$http://cxc.harvard.edu/ciao4.9$\rangle$.} with the calibration database, CALDB version 4.7.5.1\footnote{$\langle$http://cxc.harvard.edu/caldb/downloads/Release\_notes/CALDB\_v4.7.5.1.html$\rangle$.}.
The exposure times after the screening process are listed in table~\ref{observations}.
For better statistics, we combined data from the five NW observations.

\begin{figure*}
 \begin{center}
  \includegraphics[width=160mm]{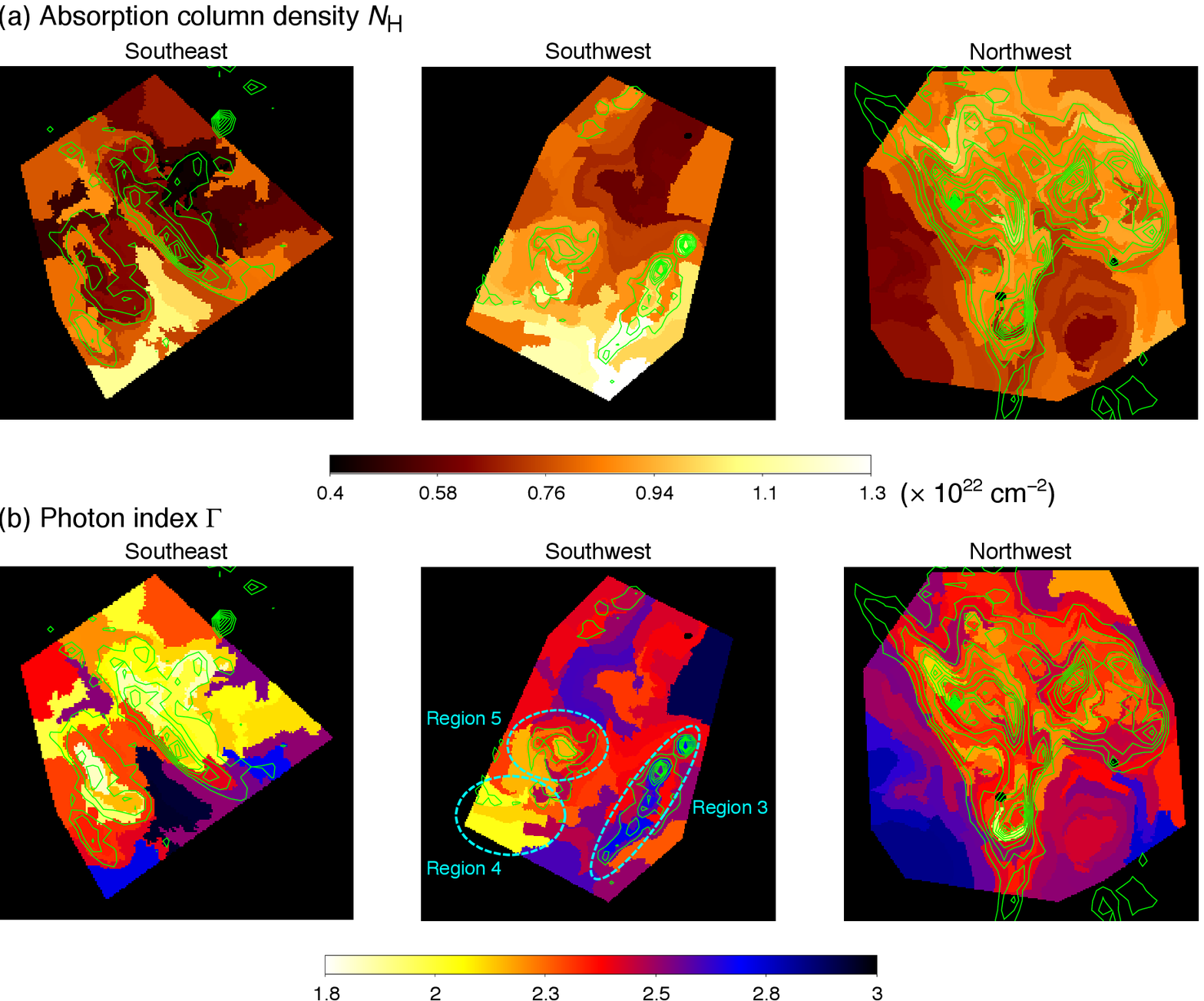}
 \end{center}
 \caption{Detailed spatial distributions of (a) the absorption column density $N_{\rm H}$ and (b) the photon index $\Gamma$. Colors indicate values, and overlaid green contours represent Chandra X-ray brightness (0.7--7~keV). Regions 3, 4, and 5 surrounded with cyan dash lines are characteristic regions (see text).}\label{mapintensity}
\end{figure*}

\begin{figure}
 \begin{center}
  \includegraphics[width=80mm]{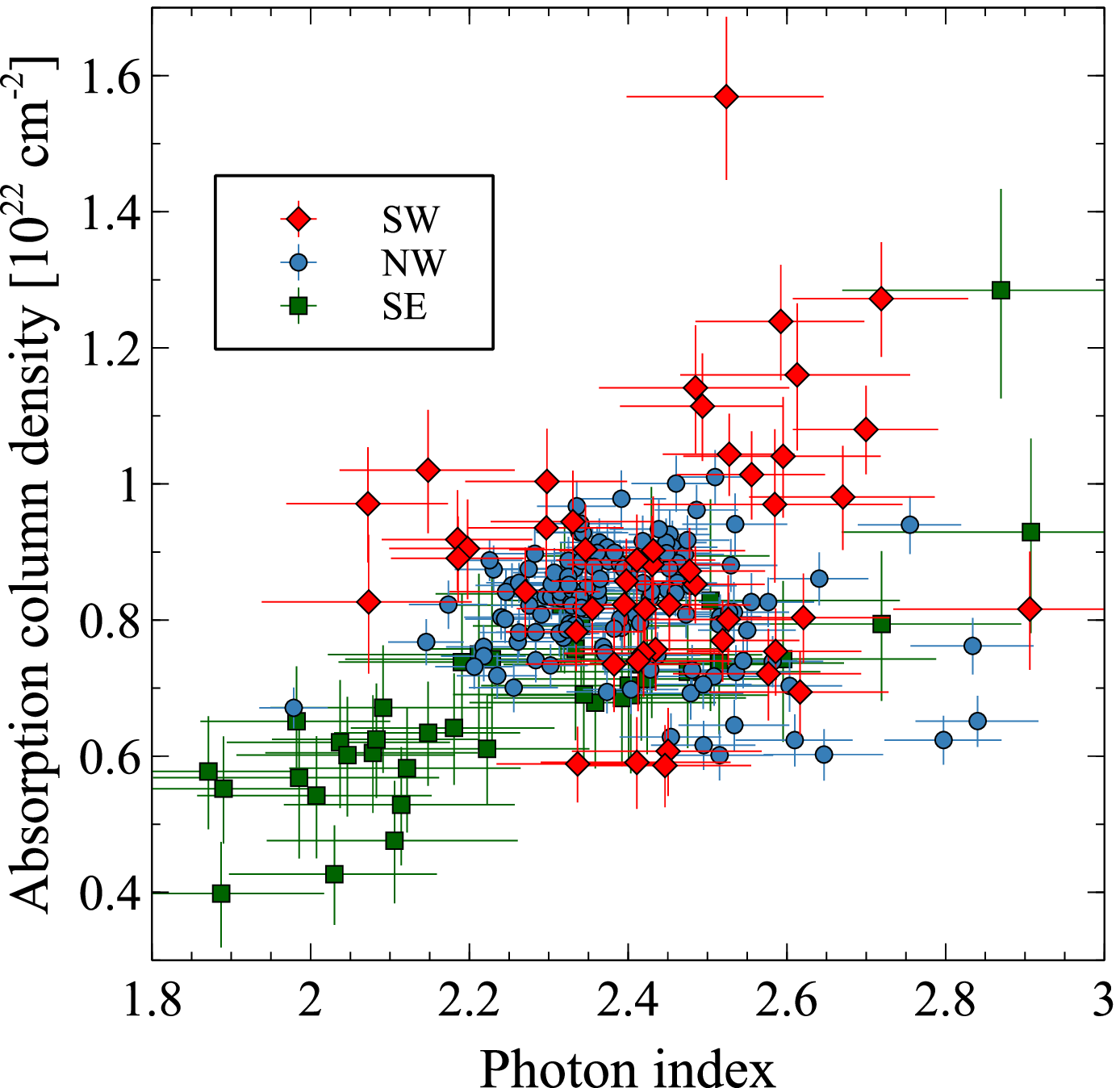}
 \end{center}
 \caption{Correlation plot between the photon index $\Gamma$ and the absorption column density $N_{\rm H}$.
Red diamonds, blue circles, and green squares correspond to the SW, NW, and SE plots, respectively.
}
\label{plot}
\end{figure}

\section{Analysis and results}\label{analysisresults}
\subsection{Imaging and spectral analysis}\label{analysis}
Figure~\ref{expcorimg} shows the 0.7--7~keV exposure-corrected images of the SE, southwestern (SW), and NW parts of RX~J1713.7$-$3946 observed with Chandra. 
Overlaid are the source- and background-extraction regions.  
The backgrounds were extracted from the off-source regions in the FOV of each observation after removing point sources found by the \texttt{wavdetect} algorithm.
From the source regions, we manually excluded some bright point sources because it is difficult to distinguish point sources from the structure of the SNR with the algorithm.
We divided the source regions into many smaller subregions by using the contour-binning algorithm, \texttt{contbin} \citep{Sanders2006}, 
which generates a set of subregions along the structure of the surface brightness so that each subregion has roughly the same signal-to-noise ratio, or about the same number of photons.
As a result, we extracted spectra from 41, 49, and 125 subregions in the SE, SW, and NW parts, respectively.
The typical angular scale of subregions are several $10\arcsec$ which corresponds to a physical scale of sub-pc at a distance of $\sim 1~\mathrm{kpc}$.
The minimum photon counts of the SE, SW, and NW subregions are 1174, 992, and 6615 counts, respectively. 
The signal to noise ratios are 1.4, 8.0, and 8.9.
The NW subregions have more photons because we combined the five observations.

We then performed spectral fittings with XSPEC version 12.9.1, using Cash statistics \citep{Cash1979} as the fitting statistics.
We obtained featureless spectra from all the subregions as many previous studies reported (e.g., \cite{Koyama1997,Slane1999,Tanaka2008}). 
Note that the region where \citet{Katsuda2015} discovered thermal emission is not covered by our analysis. 
Therefore, we adopted a power-law model attenuated by interstellar absorption, 
assuming the photoelectric absorption cross sections by \citet{Balucinska1992} and the solar abundance by \citet{Anders1989}.
The spectra were all fitted well with the model.
Examples of the spectral fittings are presented in figure~\ref{examplespec}, where we plot the hardest spectrum extracted from region 1 and the softest spectrum from region 2. 
The best-fit parameters of the former spectrum are $\Gamma = 1.87^{+0.20}_{-0.21}$, $N_{\rm H} = 0.58^{+0.13}_{-0.14} \times 10^{22}~\mathrm{cm^{-2}}$, and $F_{3\textrm{--}10~\mathrm{keV}} = (0.23 \pm 0.03) \times 10^{-12}~\mathrm{erg~s^{-1}~cm^{-2}}$. 
Fitting the latter spectrum yields $\Gamma = 2.95^{+0.36}_{-0.40}$, $N_{\rm H} = 1.06^{+0.23}_{-0.27} \times 10^{22}~\mathrm{cm^{-2}}$, and $F_{3\textrm{--}10~\mathrm{keV}} = (0.09 \pm 0.02) \times 10^{-12}~\mathrm{erg~s^{-1}~cm^{-2}}$.

\subsection{Detailed spatial distributions of the absorption column density and the photon index}\label{results}
Figure~\ref{mapintensity}a shows absorption column density maps. 
The absorption column density $N_{\rm H}$ ranges from $0.39 \times 10^{22}~\mathrm{cm^{-2}}$ to $1.6 \times 10^{22}~\mathrm{cm^{-2}}$ with the average relative error of $12\%$.
The SW and NW parts have  generally larger $N_{\rm H}$ ($\sim 0.9 \times 10^{22}~\mathrm{cm^{-2}}$) than the SE part ($N_{\rm H} \sim 0.7 \times 10^{22}~\mathrm{cm^{-2}}$).
In particular, the southern region of the SW part has very large values ($N_{\rm H} > 1.0 \times 10^{22}~\mathrm{cm^{-2}}$).
The global distribution and the range of $N_{\rm H}$ are consistent with the previous studies such as the one by  \citet{Sano2015}.

Figure~\ref{mapintensity}b shows photon index maps. 
The average relative error is $6.0\%$.
The SE part is generally harder than the western parts.
The inner region of the NW part is relatively soft ($\Gamma \sim 2.7$).
These global distribution and the parameter range are also consistent with the result by \citet{Sano2015}.
Focusing on the detailed distributions, we found that the clear filamentary regions have smaller values ($\Gamma \sim 2.0$) in the SE part.
However, the hardest regions in the western do not necessarily coincide with the brightest parts.
In the SW part, it is notable that the inner hard region and the bright region (regions 4 and 5, respectively) do not coincide with each other, whereas the outer bright filamentary region (region 3) is obviously soft.

\begin{figure*}
 \begin{center}
  \includegraphics[width=160mm]{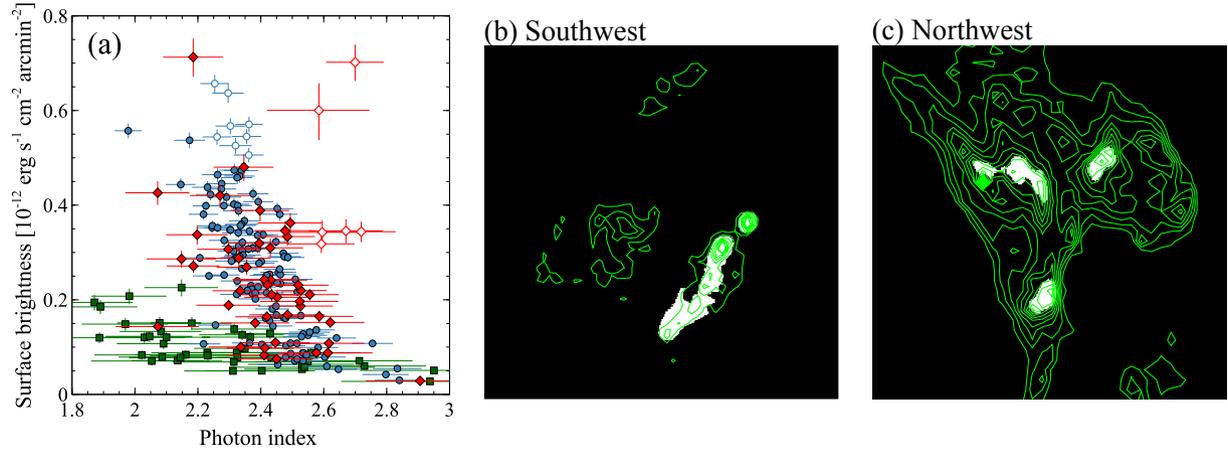}
 \end{center}
 \caption{(a) Correlation plot between the photon index $\Gamma$ and the X-ray surface brightness $\Sigma_{3\textrm{--}10~\mathrm{keV}}$. 
Symbols are the same as figure~\ref{plot} but outliers are indicated as outlined symbols. (b) Subregions corresponding to the outliers in the SW part. Overlaid contours are the same as figure~\ref{mapintensity}. (c) Same as (b) but in the NW part.}
\label{outliermap}
\end{figure*}

\begin{figure*}
 \begin{center}
  \includegraphics[width=160mm]{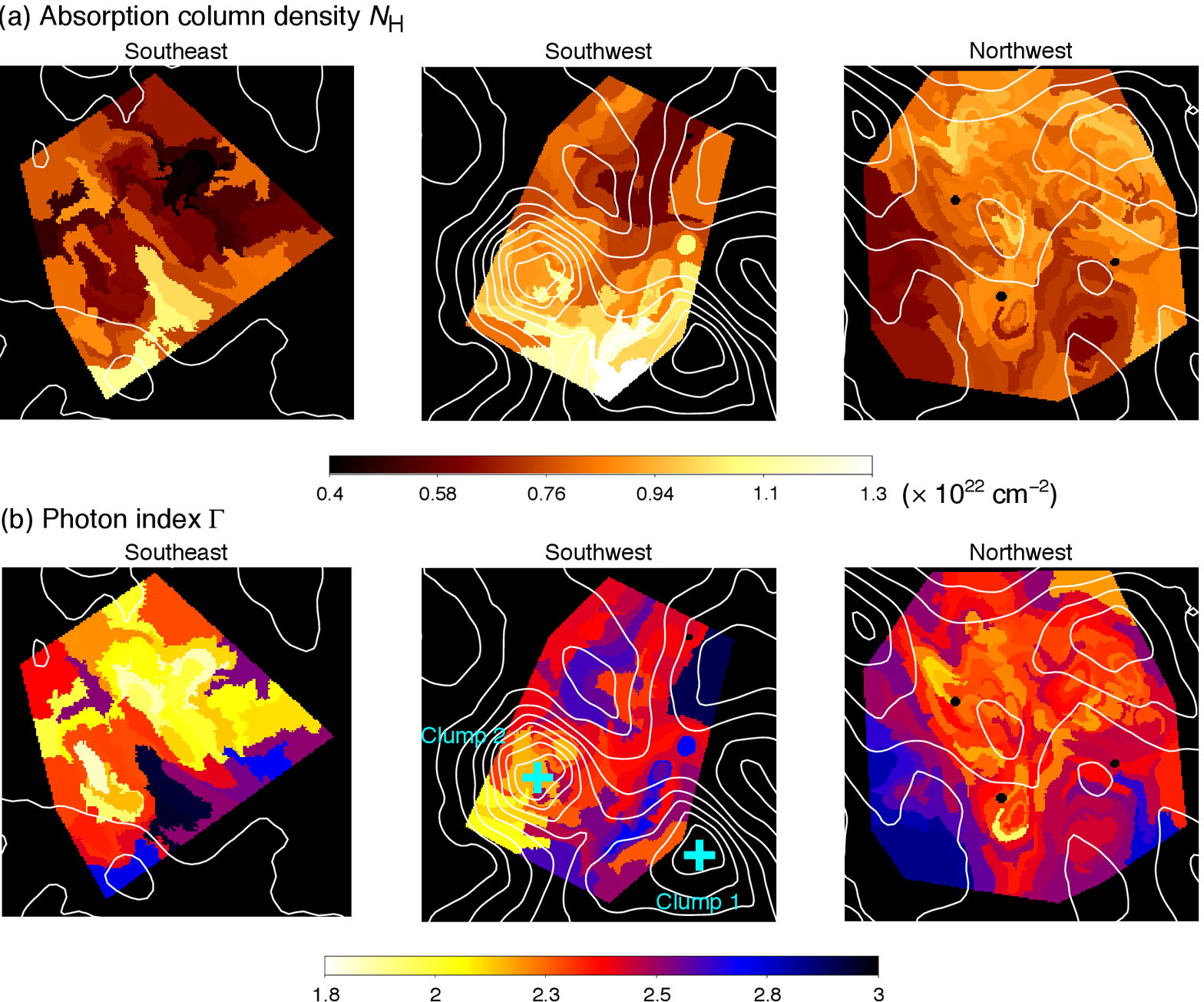}
 \end{center}
 \caption{Same maps as figure~\ref{mapintensity}, but overlaid white contours are the same as figure~\ref{wholeimage}. Two cyan crosses indicate peaks of interstellar proton density (clumps 1 and 2).}\label{mapradio}
\end{figure*}

\section{Discussion}\label{discussion}
We found significant spatial variation of the photon index in section~\ref{analysisresults}. 
However, it is of importance to check the observed correlation between the photon index and the absorption column density because they are statistically coupled when fitting with an absorbed power-law model.
As shown in figure~\ref{plot}, a very strong correlation with a  correlation coefficient $R = 0.87$ is observed in the SE part, but not in the western parts.
Therefore, we investigated the influence of the statistical couple only in the SE part.
We fixed $N_{\rm H}$ at $0.67 \times 10^{22}~\mathrm{cm^{-2}}$, which is the best-fit value for the spectrum extracted from the whole SE source region, and confirmed that the tendency for the bright regions to be hard does not change.

In the SE part, the anti-correlation between the photon index and the surface brightness is observed (figure~\ref{outliermap}a) with $R = -0.64$. 
The bright filamentary structures in SE would be associated with the location of the blast wave shock front. 
In some young SNRs such as Vela Jr. \citep{Kishishita2013}, SN~1006 \citep{Rothenflug2004,Katsuda2010}, and Tycho \citep{Cassam-Chenai2007}, the authors found a similar tendency 
that synchrotron X-rays are the hardest at the shock front with gradual steepening toward downstream. 
Our finding in the SE part would be another addition to those observational facts, and can be interpreted as a results of  the decrease of the cutoff energy of accelerated electrons 
due to synchrotron cooling (e.g., \cite{Kishishita2013}).

Contrary to the SE part, the anti-correlation between the photon index and the surface brightness in SW is weak with $R = -0.16$ with notable outliers 
plotted as open red diamonds in figure~\ref{outliermap}a. 
The outliers are all from region~3, the bright filament located close to the SNR rim which would correspond to the blast wave shock front as illustrated in figure~\ref{outliermap}b, 
and have significantly larger photon indices than the surrounding regions. 
Considering the fact that synchrotron X-rays are emitted by electrons in the cutoff region, we can regard a photon index as a proxy for a synchrotron cutoff energy, and thus the larger photon indices indicate lower synchrotron cutoff energies and vice versa \citep{Tanaka2008}.
If particle acceleration proceeds in the Bohm diffusion regime and the energy losses of electrons are dominated by synchrotron cooling, the synchrotron spectrum has a form of
\begin{equation}
\frac{dn}{d\varepsilon} \propto \varepsilon^{-2} \left[ 1+ 0.46 \left( \frac{\varepsilon}{\varepsilon_0}\right)^{0.6} \right]^{2.29} \exp \left[ -\left( \frac{\varepsilon}{\varepsilon_0} \right)^{1/2} \right]
\end{equation}
with the cutoff energy $\varepsilon_0$ given by
\begin{equation}
\varepsilon_0 = 0.55 \left( \frac{v_{\rm sh}}{3000~{\rm km}~{\rm s}^{-1}} \right)^2 \eta^{-1}~{\rm keV},  \label{ZA07}
\end{equation}
where $v_{\rm sh}$ is the shock speed and $\eta~(\ge1)$ is the so-called ``gyrofactor'' \citep{Zirakashvili2007}. 
We note that a molecular clump (clump 1 in figure~\ref{mapradio}b; see also \cite{Sano2013}, \yearcite{Sano2015}) is located just outside the filament. 
Therefore, a natural explanation of the softer spectra would be deceleration of the shock caused by interaction with the clump. 
The shock speed of the NW shell is measured to be $3900~{\rm km}~{\rm s}^{-1}$ by \citet{Tsuji2016}. 
Using this value, and comparing photon indices between the SW and NW shells, we found that the spectra of region 3 can be explained if the shock speed is $2800~{\rm km}~{\rm s}^{-1}$. 
We here assumed the gyrofactor $\eta$ is common between the two regions. 
This value is substantially smaller than $\sim 3900~{\rm km}~{\rm s}^{-1}$ in NW measured by \citet{Tsuji2016} and $\sim 3500~{\rm km}~{\rm s}^{-1}$ in SE by \citet{Acero2017}. 

Our spatially resolved spectroscopy revealed hard spectra in the inner region (region 4) in SW.  
In contrast to the hard filament in SE, the emission in region 4 does not appear to be the shock front, suggesting that additional acceleration is working such as 
acceleration at a reverse shock or a reflection shock as a result of interaction between the forward shock and dense gas. 
Such a scenario is actually discussed by \citet{Sato2018} on Cassiopeia A to account for an intense hard emission detected spatially coincident with inward-moving shocks. 
In this sense, it is worth pointing out that region 4 is located close to a molecular clump, clump 2 in figure~\ref{mapradio}. 
 \citet{Inoue2012} investigated properties of a blast wave shock interacting with clumpy interstellar gas with three-dimensional magnetohydrodynamic simulations. 
 They claimed that, if the SNR is young with a shock speed of $v_{\rm sh} \gtrsim 1000~{\rm km}~{\rm s}^{-1}$ like RX~J1713.7$-$3946 \citep{Tsuji2016}, the shock 
 generates a turbulent shell, which amplifies the magnetic field up to $1~{\rm mG}$. 
 They also predicted that a reflection shock is generated as a result of the shock-cloud interaction, and that it accelerates particles in the downstream region 
 where the magnetic field is amplified. 
Although it is difficult to confirm only with the present Chandra data, particle acceleration in a reflection shock with an amplified magnetic field is a possible 
explanation for the hard emission in region 4. 

In NW, the relation between the photon index and the surface brightness plotted in figure~\ref{outliermap}a shows relatively a strong anti-correlation with $R = -0.71$. 
However, data points extracted from bright filamentary structures again constitute outliers, similarly to SW (see figure~\ref{outliermap}c). 
We found that the outliers are in between the dense molecular clouds (clumps D and L; \cite{Sano2015}).
With Suzaku, \citet{Sano2015} reported that the spectrum of NW becomes harder toward these clumps in a sub-parsec scale.
The sub-parsec scale coincides with the scale size discussed by \citet{Uchiyama2007} and \citet{Inoue2012}.
Our Chandra analysis reveals a more detailed structure in which the hardest regions are locally distributed among the dense clumps.
It is also notable that the outliers are close to (but not coincident with) the filaments where the year-scale time variability was reported by \citet{Uchiyama2007}.
These results may  be interpreted as a consequence of shock deceleration and acceleration at a reflected shock.
If so, the magnetic field amplification evidenced by the short-time variability of X-ray flux in NW \citep{Uchiyama2007} may be caused by the shock-cloud interaction as discussed by \citet{Inoue2012}.

\section{Conclusions}\label{conclusions}
In this paper, we performed spectral analysis with Chandra archive data of RX~J1713.7$-$3946 in 2005--2011 in order to clarify X-ray properties of the filamentary structures and their surrounding regions, and compared them with the distributions of ISM.
We obtained absorption column density $N_{\rm H}$ and photon index $\Gamma$ maps in the SE, SW and NW parts of RX~J1713.7$-$3946 with an angular scale of $10\arcsec$, and found the photon index varies significantly from region to region.
Our conclusions are summarized below:

\begin{enumerate}

\item
$N_{\rm H}$ and $\Gamma$ ranges from $0.39 \times 10^{22}~\mathrm{cm^{-2}}$ to $1.6 \times 10^{22}~\mathrm{cm^{-2}}$ and from $1.8$ to $3$, respectively. 
In the absorption column density maps, the SW and NW parts have generally larger $N_{\rm H}$ than the SE part, and the southern region of the SW part yields high $N_{\rm H}$ of $> 1.0 \times 10^{22}~\mathrm{cm^{-2}}$.
In the photon index maps, the SW and NW part is softer than the SE part, and the inner region of the NW part is relatively soft ($\Gamma \sim 2.7$).
These distributions of $N_{\rm H}$ and $\Gamma$ are consistent with those given by \citet{Sano2015} with Suzaku.

\item
In the SE part, the bright filaments spatially coincide with the relatively hard regions as observed in some other young SNRs.
This can interpreted as a consequence of the decreasing cutoff energy of accelerated electrons due to synchrotron cooling.

\item
Contrary to the SE filaments, region 3 in the SW part seems to be the blast wave shock front, but is softer than surrounding regions.
This region constitutes the outliers in the correlation plot between the photon index and the surface brightness.
This most possibly reflects the deceleration of the shock wave due to interaction with the clumpy ISM (clump 1) lowers the acceleration efficiency.

\item
We also found the hard emission in the inner region (region 4) in SW, but this region does not appear to be the shock front, unlike the SE filaments.
This implies an additional particle acceleration mechanism such as the acceleration at a reverse shock or a reflected shock due to the interaction between the forward shock and dense ISM.
In fact, region 4 in SW is close to a peak of the ISM (clump 2).

\item
In NW, we also found some outliers in the correlation plot between the photon index and the surface brightness.
It is known that the NW part mainly interacts with dense molecular gas \citep{Fukui2012}, and we first found the sub-pc anti-correlation between hard region and dense ISM, similarily to SW.
Therefore, the hardness of spectra may be determined based on a balance of shock deceleration and acceleration at a reflected shock.

\end{enumerate}

\begin{ack}
We thank Dr. Koji Mori for providing us with the XMM-Newton image.
We are also grateful to Dr. Hidetoshi Sano for providing us with the NANTEN, and ATCA \& Parkes data.
This work is supported by JSPS/MEXT Scientific Research Grant Numbers JP25109004 (T.T. and T.G.T.), JP26800102 (H.U.), JP15J01842 (H.M.), and JP15H02090 (T.G.T.).
\end{ack}

\end{document}